\documentclass[10pt,amsmath,amssymb,amsfont,aps,prd,twocolumn,nofootinbib,superscriptaddress]{revtex4-2}
\usepackage[utf8]{inputenc}
\usepackage[english]{babel}
\usepackage[T1]{fontenc}
\usepackage[colorlinks,citecolor=blue,linkcolor=blue,urlcolor=blue, breaklinks=true]{hyperref}
\usepackage{standalone,graphicx,pgf,tikz,bm,braket,url,orcidlink,bbold,lmodern}

\usetikzlibrary{arrows.meta,positioning,calc}

\usepackage{xcolor, xspace, ifthen}
\newcommand{\showcomments}{true} 

\newcommand{\ofek}[1]%
{\ifthenelse{\equal{\showcomments}{true}}%
{{\color{purple}{\small \textbf{Ofek:} #1}}}{\xspace}}%

\newcommand{\m}[1]%
{\ifthenelse{\equal{\showcomments}{true}}%
{{\color{blue}{\small \textbf{m:} #1}}}{\xspace}}%

\newcommand{\hannah}[1]%
{\ifthenelse{\equal{\showcomments}{true}}%
{{\color{orange}{\small \textbf{H:} #1}}}{\xspace}}%

\definecolor{darkgreen}{rgb}{0.0, 0.5, 0.13}

\newcommand{\tdg}[1]%
{\ifthenelse{\equal{\showcomments}{true}}%
{{\color{darkgreen}{\small \textbf{TG:} #1}}}{\xspace}}%

\newcommand{\Sc}{\mathcal{S}}
\newcommand{\Gc}{\mathcal{G}}
\newcommand{\Hc}{\mathcal{H}}

\newcommand{\inv}{\mathrm{inv}}
\newcommand{\SO}{\mathrm{SO}}
\newcommand{\SU}{\mathrm{SU}}
\newcommand{\Tr}{\mathrm{Tr}}
\newcommand{\I}{\mathbb{1}}
\newcommand{\Nl}{\mathbb{N}}
\newcommand{\Cl}{\mathbb{C}}

\newcommand{\ketbra}[2] {| #1 \rangle \! \langle #2 |}

\begin{document}
\title{Complete Relational Description of Spin in a Quantum Background}
\date{\today}
\author{Hannah Troger\;\orcidlink{0009-0002-3489-7023}}
\thanks{These authors contributed equally to this work.}
\affiliation{Institute for Quantum Optics and Quantum Information (IQOQI) Vienna, Austrian Academy of Sciences, Boltzmanngasse 3, A-1090 Vienna, Austria}
\affiliation{University of Vienna, Faculty of Physics, Vienna Center for Quantum Science and Technology (VCQ), Boltzmanngasse 5, A-1090 Vienna, Austria}
\author{Ofek Bengyat\;\orcidlink{0000-0001-5547-9176}}
\thanks{These authors contributed equally to this work.}
\affiliation{Institute for Quantum Optics and Quantum Information (IQOQI) Vienna, Austrian Academy of Sciences, Boltzmanngasse 3, A-1090 Vienna, Austria}
\affiliation{University of Vienna, Faculty of Physics, Vienna Center for Quantum Science and Technology (VCQ), Boltzmanngasse 5, A-1090 Vienna, Austria}
\author{Thomas D. Galley\;\orcidlink{0000-0002-8870-3215}}
\affiliation{Institute for Quantum Optics and Quantum Information (IQOQI) Vienna, Austrian Academy of Sciences, Boltzmanngasse 3, A-1090 Vienna, Austria}

\author{Marios Christodoulou\;\orcidlink{0000-0001-6818-2478}}
\affiliation{Institute for Quantum Optics and Quantum Information (IQOQI) Vienna, Austrian Academy of Sciences, Boltzmanngasse 3, A-1090 Vienna, Austria}

\begin{abstract}
The standard description of the state of a spin in quantum mechanics presupposes externally fixed directions---a classical background. Can a spin be fully described instead in relation to other quantum mechanical systems? Poulin suggested twenty years ago group averaging over rotations the joint state of a fundamental spin and a reference spin with large angular momentum which, however, yields a classical bit in a probabilistic mixture. We revisit this idea and show that when the quantum reference system is augmented to \emph{two} large spins, the standard quantum mechanical description of a spin is recovered in the limit of large quantum numbers for the reference system.

\end{abstract}
\maketitle

\section{Introduction}

The standard quantization of a spin makes reference to a fixed external classical frame, which defines directions along which the spin may be measured. This frame could be imagined to be physically realized, for instance, by the walls of the laboratory. More generally, this `classical background' corresponds to a space geometry, such as a Euclidean space.

However, if all physical systems obey quantum mechanics, including spacetime, the general description of a quantum system---of a spin in particular---should be in relation to other quantum systems. With a view towards a quantum theory of spacetime, such `quantum frames of reference' have been considered since the late 80s \cite{Aharonov_1984,Rovelli_1991,toller_quantum_1996,toller_quantum_1996,butterfield_aspects_1999}. The description of a system relative to a quantum frame of reference will not depend only on properties of the system of interest, as the algebra of relative observables will depend on the frame~\cite{AliAhmad_2022}. Over the past decade much effort has gone into developing a comprehensive framework for quantum reference frames e.g.~\cite{poulin_dynamics_2007,girelli_quantum_2008,angelo_kinematics_2012,Pienaar_2016, Angelo_2011,pereira_galilei_2015,Loveridge_2018,delaHamette_2020,Vanrietvelde2020changeof,Krumm_2021,delaHamette_2021,Hohn_2021,Hohn_2021a,PhysRevResearch.3.043138,Carette2025operationalquantum,Castro-Ruiz_2025,Kabel_2025,Cepollaro_2025,DeLaHamette_2025}. This has lead to an understanding, for instance, that entanglement~\cite{Giacomini_2019,Krumm_2021,Cepollaro_2025} and localization of events \cite{Kabel_2025,DeLaHamette_2025} are relative notions,  depending on the choice of frame.

Continuum spacetime is suspected to emerge from an underlying discrete, combinatorial physics, modeled as a spin network, e.g.~\cite{Penrose,Rovelli:1995ac,linnemann2018hintsa,Oriti_2021}. In this setting, a fixed background classical spacetime cannot be presupposed. A spin would have to be described against other quantized angular momenta.  With these motivations in mind, we consider here the simplest case study of a single fundamental spin, and ask: \emph{how can a quantum spin be described  in relation to a reference quantum system?}\footnote{See also \cite{PhysRevResearch.3.043138}. That work extends QRF transformations to spin systems, with the frame consisting of infinite spins from the outset. This is unlike what we do here, as (i) a frame of infinite spins is the classical limit of our construction (see Section \ref{sec:clasLimit}) and (ii) we do not consider transformations but rather the extraction of frame-independent information.}

We had to overcome two difficulties. At first glance, it seems intuitive that a spin can be fully described in relation to any other physical system that `points' towards a direction. But, this is only sufficient for a classical spin. For a quantum spin, it is not enough to relationally encode that the joint state is a (mostly) aligned or a (mostly) anti-aligned state. It is also necessary to encode complementarity: the system of reference needs to be a composite system, with two angular momenta that do not commute. This leads to the second difficulty, which is technical: the resulting Clebsch–Gordan coefficients are somewhat complicated.

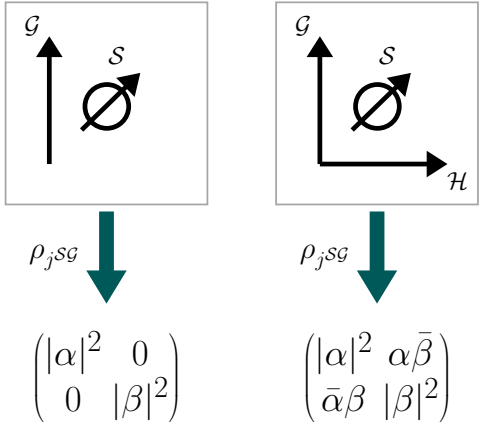
\begin{figure}[t]
    \centering
    \begin{tikzpicture}[
  topbox/.style={draw=gray!70, very thick, fill=white, minimum width=6.0cm, minimum height=6.0cm, inner sep=0pt},
  nobox/.style={draw=none, fill=none, minimum width=6.0cm, minimum height=3.2cm, inner sep=0pt},
  downarr/.style={-{Triangle[length=8mm,width=9mm]}, line width=10pt, draw=teal!70!black},
  axis/.style={-{Triangle[length=5mm,width=6mm]}, line width=3.2pt, draw=black},
  bitstroke/.style={line width=3.2pt, draw=black},
  bitarrow/.style={-{Triangle[length=5mm,width=6mm]}, line width=3.2pt, draw=black},
  every node/.style={font=\large}
]

\def\TopScale{0.8}

\def\AxisLen{3.80cm}

\def\MatSize{\Huge}

\begin{scope}[scale=\TopScale, transform shape]
  \node[topbox] (TL) {};
  \node[topbox, right=2.0cm of TL] (TR) {};

  \begin{scope}
    \coordinate (O) at ($(TL.center)+(-1.7cm,-1.8cm)$);

    \draw[axis] (O) -- ($(O)+(0,\AxisLen)$);     

    \node at ($(O)+(-0.5cm,\AxisLen+0.3cm)$) {\huge$\mathcal{G}$};

    \coordinate (cTR) at ($(O)+(0.45*\AxisLen,0.45*\AxisLen)$);
    \draw[bitstroke] (cTR) circle [radius=0.65cm];
    \draw[bitarrow]  ($(cTR)+(-0.75cm,-0.7cm)$) -- ($(cTR)+(1cm,1cm)$);
    \node at ($(cTR)+(0.3cm,1.5cm)$) {\huge$\mathcal{S}$};
  \end{scope}

  \begin{scope}
    \coordinate (O) at ($(TR.center)+(-1.7cm,-1.8cm)$);

    \draw[axis] (O) -- ($(O)+(0,\AxisLen)$);     
    \draw[axis] (O) -- ($(O)+(\AxisLen,0)$);    

    \node at ($(O)+(-0.5cm,\AxisLen+0.3cm)$) {\huge$\mathcal{G}$};
    \node at ($(O)+(\AxisLen+0.3cm,-0.5cm)$) {\huge$\mathcal{H}$};

    \coordinate (cTR) at ($(O)+(0.45*\AxisLen,0.45*\AxisLen)$);
    \draw[bitstroke] (cTR) circle [radius=0.65cm];
    \draw[bitarrow]  ($(cTR)+(-0.75cm,-0.7cm)$) -- ($(cTR)+(1cm,1cm)$);
    \node at ($(cTR)+(0.3cm,1.5cm)$) {\huge$\mathcal{S}$};
  \end{scope}

  \coordinate (TopAnchorBL) at (TL.south);
  \coordinate (TopAnchorBR) at (TR.south);
\end{scope}

\node[nobox, below=2.5cm of TopAnchorBL] (BL) {};
\node[nobox, below=2.5cm of TopAnchorBR] (BR) {};

\draw[downarr]
  ($(TopAnchorBL)+(0,-0.15)$) -- ($(BL.north)+(0,0.15)$)
  node[midway, left=10pt] {\huge$\rho_{j^{\mathcal{SG}}}$};

\draw[downarr]
  ($(TopAnchorBR)+(0,-0.15)$) -- ($(BR.north)+(0,0.15)$)
  node[midway, left=10pt] {\huge$\rho_{j^{\mathcal{SG}}}$};

\node at (BL.center) {\MatSize$\begin{pmatrix}
|\alpha|^{2} & 0\\
0 & |\beta|^{2}
\end{pmatrix}$};

\node at (BR.center) {\MatSize$\begin{pmatrix}
|\alpha|^{2} & \alpha\bar{\beta}\\
\bar{\alpha}\beta & |\beta|^{2}
\end{pmatrix}$};

\end{tikzpicture}
    \caption{A spin-1/2 system $\mathcal{S}$ alongside a large spin system $\Gc$ yields a relational state approximately corresponding to a mixture, see Section~\ref{sec:one_directional reference}. Adding another large spin system $\Hc$ to the reference, yields a relational state which approaches an exact encoding of the qubit $\Sc$ as the quantum numbers of the reference grow, see Section~\ref{sec:two_directional references}. Note that analogous states may be written for $\rho_{j^\mathcal{SH}}$.}
    \label{fig:first} 
\end{figure}

The calculation presented in this work builds on an idea by Poulin from 2006 \cite{Poulin_2006}. They considered the joint state of a spin-1/2 system with a second, larger spin, removing the fiducial information of the coordinate axes by averaging the density matrix over rotations. However, this yields a mixed state for the spin, no different than the state of a classical probabilistic bit (see Section \ref{sec:one_directional reference}). We will show that augmenting the quantum reference frame with a second large spin and following a similar procedure, yields a relational state that includes information on the coherence of the spin-1/2 system. The encoding becomes approximately exact and the state of the spin-1/2 becomes approximately pure when quantum numbers for the reference are taken large (see Section \ref{sec:two_directional references}). This is our main result, sketched in Figure \ref{fig:first}.

In Section \ref{sec:comments}, we discuss: (i) the formal classical limit of infinite angular momenta for the reference (Section \ref{sec:clasLimit}), (ii) the conceptual difference between incoherent and coherent average, in particular, that complete removal of the background from the relational encoding requires the use of the latter, which in turn requires a different reference state (Section \ref{sec:coherent}), and (iii) demonstrate that the general morals learned carry through for systems of interest with many degrees of freedom (Section \ref{sec:frameQuantisation}).

\section{Spin in relation to one reference spin}
\label{sec:one_directional reference}
Poulin discussed the following strategy to extract a relational description of a quantum spin \cite{Poulin_2006}. Start from the standard quantum mechanical description of a joint system consisting of a spin-1/2, which is the system of interest, and a reference spin. The dependence on classical externally defined directions is removed by (incoherently, see below) group averaging the joint state over rotations. Finally, take the limit of large angular momentum for the reference spin.  

Let us illustrate the above concretely. Consider the spin-$1/2$ state
\begin{equation}
\label{eq:spin state}
    \ket{\psi}^\Sc = \alpha \ket{\uparrow}^\Sc_z + \beta \ket{\downarrow}^\Sc_z.
\end{equation}
The $z$-axis can be physically modeled as a second quantum system $\mathcal{G}$ with angular momentum $\hat{J}^\mathcal{G}$, prepared in its highest magnetic moment state $\ket{G, G}^\Gc_z$. That is, we have that $ \hat{M}_z^\mathcal{G} \ket{G, G}^\Gc_z = G \ket{G, G}^\Gc_z  $, where $\hat{M}^\mathcal{G}_z$ is the magnetic moment of $\mathcal{G}$ in the $z$-direction. The state $\ket{G, G}^\Gc_z$ is chosen because it is the `most' semi-classical among the states $\ket{G, M}^\Gc_z$, in the sense that it saturates the Heisenberg uncertainty relation. Denoting as $\Delta M_x$ the standard deviation of $\hat{M}_x$ on $\ket{G, G}^\Gc_z$, and similarly for $\Delta M_y$, we have that $\Delta M_x \Delta M_y = \frac{G}{2}$, see e.g~\cite{perelomov_generalized_1986}. Intuitively, the uncertainty cone around the most probable direction of the spin is minimal.

The spin system $\Sc$ and the reference system $\Gc$ do not interact with each other. In the total angular momentum eigenbasis, their joint state is
\begin{align}
\label{al:psiSG}
    \ket{\psi}^{\Sc\Gc} &= \left(\alpha \ket{\uparrow}^\Sc_z + \beta \ket{\downarrow}^\Sc_z \right) \otimes \ket{G,G}^\Gc_z \\\nonumber
    &= \alpha \ket{G+\tfrac{1}{2}, G+\tfrac{1}{2}}_z^{\Sc\Gc} \\\nonumber 
    &+ \frac{\beta}{\sqrt{2G+1}} \ket{G+\tfrac{1}{2}, G-\tfrac{1}{2}}_z^{\Sc\Gc} \\\nonumber
    &+ \frac{\beta \sqrt{2G}}{\sqrt{2G+1}} \ket{G-\tfrac{1}{2}, G-\tfrac{1}{2}}_z^{\Sc\Gc}
\end{align}
The prefactors are the Clebsch-Gordan (CG) coefficients for coupling a spin-$1/2$ with a spin $G$.

The dependence on the external $z$-axis can be removed by group averaging the density matrix $\rho^{\Sc\Gc} = \ketbra{\psi}{\psi}^{\Sc\Gc}$ over rotations $\omega \in SO(3)$:
\begin{equation}\label{eq:incoherent_average_sg}
    \mathcal{E}(\rho^{\Sc\Gc}) = \int_{SO(3)} \!\!\!\text{d}\omega \; R^{\Sc\Gc}(\omega) \;\rho\;  R^{\Sc\Gc}(\omega)^\dag,
\end{equation}
where $\text{d}\omega$ denotes the Haar measure on $SO(3)$. The above group averaging may be called \emph{incoherent} because it acts diagonally on the density matrix $\rho^{\Sc\Gc}$. However, we will see that this form of group averaging does \emph{not} mean that the spin cannot have quantum coherence; rather, it removes coherence between sectors of different \emph{total} angular momentum (see the calculation in the next Section and the discussion in Section~\ref{sec:clasLimit}).

The representation $R^{\Sc\Gc}(\omega)$ of the rotation group acts on the systems $\Sc$ and $\Gc$ as
\begin{equation}
    R^{\Sc\Gc}(\omega) = D^{1/2}(\omega) \otimes D^G(\omega),
\end{equation}

\noindent where $D^j(\omega)$ is the Wigner rotation matrix in the spin-$j$ representation. Note that for $j=1/2$, this is defined only up to a sign which cancels under conjugation and leaves the density operator unchanged. The resulting state is 
\begin{align}
\label{eq:step1}
    \mathcal{E}(\rho^{\Sc\Gc}) = &\left(|\alpha|^2 + \frac{|\beta|^2}{2G+1}\right) \ketbra{G+\tfrac12}{G+\tfrac12} \otimes \frac{\mathbb{1}_{2G+2}}{2G+2} \\\nonumber
    &+ \frac{ 2G |\beta|^2}{2G+1}  \ketbra{G-\tfrac12}{G-\tfrac12}  \otimes \frac{\mathbb{1}_{2G}}{2G},
\end{align}
where $\ketbra{G\pm\frac12}{G\pm\frac12}$ are projectors on the corresponding total angular momentum $j^{\Sc\Gc}$ eigenspace. The above state is diagonal in $j^{\Sc\Gc}$, due to the diagonal group averaging. The sector of the magnetic component along $z$ is now maximally mixed: the information for $\hat{M}^\mathcal{G}_z$, which is defined with respect to the fiducial background $z$-axis, has been erased. Tracing out the magnetic moment sector, we get the state
\begin{align}
\label{eq:step2}
    \tilde{\mathcal{E}}(\rho^{\Sc\Gc}) = &|\alpha|^2 \ketbra{G+\tfrac12}{G+\tfrac12} \nonumber \\ 
    &+ \frac{|\beta|^2}{2G+1} \ketbra{G+\tfrac12}{G+\tfrac12} \nonumber \\
    &+ \frac{2G|\beta|^2}{2G+1}  \ketbra{G-\tfrac12}{G-\tfrac12}.
\end{align}
Let us pause to make a comment of conceptual relevance. As expected, the above state makes no reference to a system of axes: the two states $G \pm \tfrac12$ correspond to the two systems being aligned or anti-aligned, a relational statement. However, it raises the question: how to interpret the fact that the joint system has \emph{total} angular momentum $G \pm \tfrac12$? This seems to imply that both systems are rotating with respect to some other, background system. This implicit presence of background can be also removed via the use of a coherent average, see Section \ref{sec:coherent}.

Let us now inspect the limit when the quantum number of the reference is large. When $G \gg 1$ the second term in \eqref{eq:step2} is negligible, and the state is approximately
\begin{align}\label{eq:rhophys}
     \tilde{\mathcal{E}}^C(\rho^{\Sc\Gc}) \approx |\alpha|^2  \ketbra{G+\tfrac12}{G+\tfrac12}
    + |\beta|^2  \ketbra{G-\tfrac12}{G-\tfrac12}.
\end{align}
This state is the probabilistic mixture of a classical spin with probability $|\alpha|^2$ of being in the spin-up state and probability $|\beta|^2$ of being in the spin-down state. Note that the above result can be seen directly from \eqref{al:psiSG} and the fact that \eqref{eq:step1} projects on eigenspaces of the total angular momentum of $\Sc\Gc$. We have presented the intermediate steps in this simpler setting, because in the following Section they become necessary. 

To understand the apparent loss of coherence, it is instructive to consider as a starting point (instead of the pure state \eqref{eq:spin state}) the probabilistic mixture
\begin{align}
    \mu^\Sc = |\alpha|^2 \ketbra{\uparrow}{\uparrow}^\Sc+|\beta|^2 \ketbra{\downarrow}{\downarrow}^\Sc.
\end{align}
The joint state of $\mathcal{S}$ and $\Gc$ becomes
\begin{align}
    \mu^{\Sc\Gc} = \; &|\alpha|^2 \ketbra{\uparrow}{\uparrow}^\Sc \ketbra{G,G}{G,G}^\Gc \\\nonumber
    + &|\beta|^2 \ketbra{\downarrow}{\downarrow}^\Sc\ketbra{G,G}{G,G}^\Gc.
\end{align}
In the total angular momentum eigenbasis, it reads
\begin{align}
    \mu^{\Sc\Gc} = \; &|\alpha|^2 \ketbra{G+\tfrac12, G+\tfrac12}{G+\tfrac12, G+\tfrac12}^{\Sc\Gc}_z \\\nonumber
    + &\frac{|\beta|^2}{2G+1} \ketbra{G+\tfrac12, G-\tfrac12}{G+\tfrac12, G-\tfrac12}^{\Sc\Gc}_z \\\nonumber
    + &\frac{2G|\beta|^2}{2G+1} \ketbra{G-\tfrac12, G-\tfrac12}{G-\tfrac12, G-\tfrac12}^{\Sc\Gc}_z \\\nonumber
    + &\frac{\sqrt{2G}\,|\beta|^2}{2G+1} \ketbra{G+\tfrac12, G-\tfrac12}{G-\tfrac12, G-\tfrac12}^{\Sc\Gc}_z \\\nonumber
    + &\frac{\sqrt{2G}\,|\beta|^2}{2G+1} \ketbra{G-\tfrac12, G-\tfrac12}{G+\tfrac12, G-\tfrac12}^{\Sc\Gc}_z.
\end{align}
Note that the diagonal of $\mu^{\Sc\Gc}$ is the same as that of the density matrix $\rho^{\Sc\Gc}=\ketbra{{\psi}^{\Sc\Gc}}{{\psi}^{\Sc\Gc}}$ where $\ket{\psi^{\Sc\Gc}}$ is the pure joint state \eqref{al:psiSG}. Therefore, the incoherent group average applied on  $\mu^{\Sc\Gc}$ will yield the same result as on $\rho^{\Sc\Gc}$, as it only keeps the diagonal terms in the total angular momentum $j^{\Sc\Gc}$. Then, tracing out the magnetic component sector, again returns the relational state \eqref{eq:rhophys}.

The above implies that with a choice of reference system consisting of a single large spin, only the magnitude of the spin amplitudes can be encoded in the relational state, not their relative phase. The apparent loss of coherence of the target spin-1/2 is \emph{not} due to averaging incoherently over total system rotations. As the next section shows, when the reference system is composite and made of \emph{two} large spins, the same procedure produces a complete relational encoding of the pure spin state.

\section{Spin in relation to two reference spins}\label{sec:two_directional references}

We now augment the reference system with a second spin system $\Hc$, which is in a state orthogonal to $\Gc$ with respect to the reference system of axes. That is, the point of departure is the joint state of the spin-$1/2$ system $\Sc$ with the two spins $\Gc$ and $\Hc$, taken to be in the states $\ket{G,G}^\Gc_z$ and $\ket{H,H}^\Hc_x$:
\begin{align}
    \ket{\psi}^{\Sc\Gc\Hc} = \left(\alpha \ket{\uparrow}^\Sc_z + \beta \ket{\downarrow}^\Sc_z\right) \otimes \ket{G,G}^\Gc_z \otimes \ket{H,H}^\Hc_x.
\end{align}

Since we have a composite reference system $\Gc\Hc$, in order to use recoupling theory we need to make an arbitrary choice on whether to first couple the spin system $\Sc$ to $\Gc$ or $\Hc$. That is, on whether to first consider the recoupling basis over angular momenta $j^{\Sc\Gc}$ or $j^{\Sc\Hc}$. We choose to first couple $\Sc$ to $\Gc$, and then $\Sc\Gc$  to $\Hc$. The result for when $\Sc$ is first coupled to $\Hc$ can also be surmised at the end of the calculation. The combinatorial coefficients are significantly more involved than the case presented in the previous Section where only one reference large spin was used. Below, we give the main steps of the calculation and refer to Appendix~\ref{ap:calculation} for details.

\subsection{Basis Change}

We start by changing basis to the eigenbasis of angular momentum $j^{\Sc\Gc}$ and magnetic moment $m^{\Sc\Gc}$ of the joint system of $\Sc$ and $\Gc$. This change of basis is of the form
\begin{align}
    \left\{\ket{j^\Sc, m^\Sc}\right\} \otimes  \left\{\ket{j^\Gc, m^\Gc}\right\} &\otimes  \left\{\ket{j^\Hc, m^\Hc}\right\}  \\\nonumber
     \rightarrow  \left\{\ket{j^{\Sc\Gc}, m^{\Sc\Gc}, j^\Sc, j^\Gc}\right\} &\otimes \left\{\ket{j^\Hc, m^\Hc}\right\}. 
\end{align}
For brevity, since the eigenvalues $j^\Sc = S$ and $j^\Gc = G$ are constant, we omit the corresponding state factors $\otimes \ket{j^\Sc = S}$ and $\otimes \ket{j^\Gc = G}$. That is, we use the shorthand notation
\begin{align}
    \ket{j^{\Sc\Gc}, m^{\Sc\Gc}} \equiv \ket{j^{\Sc\Gc},m^{\Sc\Gc}} \otimes \ket{j^\Sc = S} \otimes \ket{j^\Gc = G}.
\end{align}
On this basis, the state $\ket{\psi}^{\Sc\Gc\Hc}$ is given by
\begin{align}\label{eq:SGxH}
        \ket{\psi}^{\Sc\Gc\Hc} = \Big(&\alpha \ket{G+\tfrac{1}{2}, G+\tfrac{1}{2}} \\\nonumber
        + &\frac{\beta}{\sqrt{2G+1}} \ket{G+\tfrac{1}{2}, G-\tfrac{1}{2} }  \\\nonumber
        + &\frac{\beta\,\sqrt{2G}}{\sqrt{2G+1}} \ket{G-\tfrac{1}{2}, G-\tfrac{1}{2}} \Big)
        \otimes \ket{H, H}^\Hc_x.
\end{align}

Now, we change basis again to the eigenbasis that has as eigenvalues the angular momentum $j^{\Sc\Gc\Hc}$ of the joint system of $\Sc$, $\Gc$ and $\Hc$. This is of the form
\begin{align} 
      \left\{\ket{j^{\Sc\Gc}, m^{\Sc\Gc}}\right\} \otimes \left\{\ket{ j^\Hc, m^\Hc}\right\} 
      \rightarrow   \left\{\ket{j^{\Sc\Gc\Hc}, m^{\Sc\Gc\Hc}, j^{\Sc\Gc}}\right\}.
\end{align}
As with the constant $j^\Sc = S$ and $j^\Gc = G$ factors, we have also suppressed $\otimes \ket{j^\Hc = H}$ from the ket-bra notation. To implement this basis change we rewrite the state $\ket{H,H}^\Hc_x$ in the $z$-basis:
\begin{align}\label{eq:Hx=Hz}
    \ket{H, H}^{\Hc}_x  = \frac{1}{2^H} \sum_{h\,=-H}^{H} \sqrt{\binom{2H}{H+h}}  \ket{H, h}^{\Hc}_z.
\end{align}
Inserting this into \eqref{eq:SGxH} and defining  the CG coefficients
\begin{align}\label{eq:cg}
    C^{j^{\Sc\Gc\Hc} m^{\Sc\Gc\Hc}}_{j^{\Sc\Gc} m^{\Sc\Gc} m^\Hc}   :=  \braket{  j^{\Sc\Gc\Hc} m^{\Sc\Gc\Hc} j^{\Sc\Gc}   \vert   j^{\Sc\Gc} m^{\Sc\Gc} j^\Hc m^\Hc },
\end{align}
the state $\ket{\psi}^{\Sc\Gc\Hc}$ takes the form 

\begin{align}\label{eq:psiSGH}
    \ket{\psi}^{\Sc\Gc\Hc} =& \frac{1}{2^H} \sum_{h=-H}^{H} \sum_{J=G-1/2+h}^{G+1/2+H} \sqrt{\binom{2H}{H+h}}\\\nonumber
    \Bigg[ &\alpha \, C^{J, G+1/2+h}_{G+1/2,  G+1/2 , h} \ket{J, G+\tfrac{1}{2}+h, G+\tfrac{1}{2}} \\\nonumber
    + \,& \frac{\beta}{\sqrt{2G+1}}  C^{J, G-1/2+h}_{G+1/2, G-1/2,  h} \ket{J, G-\tfrac{1}{2}+h, G+\tfrac{1}{2}} \\\nonumber
    + \,& \frac{\sqrt{2G} \, \beta}{\sqrt{2G+1}}  C^{J, G-1/2+h}_{G-1/2, G-1/2 , h} \ket{J, G-\tfrac{1}{2}+h, G-\tfrac{1}{2}} \Bigg],
\end{align}
see Appendix~\ref{app:A1} for intermediate steps. We use the convention that $ C^{j^{\Sc\Gc\Hc} m^{\Sc\Gc\Hc}}_{j^{\Sc\Gc} m^{\Sc\Gc} m^\Hc} = 0$ unless
\begin{equation}
        |j^{\Sc\Gc}-H|
    \leq j^{\Sc\Gc\Hc}
    \leq j^{\Sc\Gc}+H,
    \qquad
    |m^{\Sc\Gc\Hc}|
    \leq j^{\Sc\Gc\Hc}.
\end{equation}

\subsection{Group Averaging}
To obtain the relational state, which makes no reference to a fixed system of axes,
we now calculate the incoherent group average of
the density matrix $\rho^{\Sc\Gc\Hc} =  \ketbra{\psi}{\psi}^{\Sc\Gc\Hc}$  (given explicitly in Appendix~\ref{app:A2}), corresponding to the state \eqref{eq:psiSGH}. In analogy to \eqref{eq:incoherent_average_sg}, the incoherent group averaging operation is defined as
\begin{equation}\label{eq:incoherent_average_sgh}
    \mathcal{E}(\rho^{\Sc\Gc\Hc}) = \int_{SO(3)} \!\!\! \text{d}\omega \;R^{\Sc\Gc\Hc}(\omega)\; \rho^{\Sc\Gc\Hc} \; R^{\Sc\Gc\Hc}(\omega)^\dag.
\end{equation}
After tracing out the identity on the magnetic moment sector (compare to \eqref{eq:step1} and \eqref{eq:step2} in the previous Section), we get
\begin{align}\label{eq:twirled_rhoSGH}
    \tilde{\mathcal{E}}(\rho^\mathcal{SGH}) = &\sum_{J = G-1/2-H}^{G+1/2+H}  \Bigg[  \\\nonumber
    &\frac{\left|\alpha\right|^2}{2}\,  S^{(1)}_{JGH} \ketbra{J, G+\tfrac12}{J, G+\tfrac12} \\\nonumber
    +&  \frac{\sqrt{2G}\alpha \Bar{\beta}}{\sqrt{2G+1}} S^{(2)}_{JGH} \ketbra{J, G+\tfrac12}{J, G-\tfrac12}  \\\nonumber
    +&  \frac{G \, \left|{\beta}\right|^2}{2G+1}\,  S^{(3)}_{JGH} \ketbra{J, G-\tfrac12}{J, G-\tfrac12} \\\nonumber
    +&  \frac{\sqrt{2G} \, \left|{\beta}\right|^2}{2G+1} \, S^{(4)}_{JGH} \ketbra{J, G+\tfrac12}{J, G-\tfrac12}  \\\nonumber
    +&  \frac{\alpha \Bar{\beta} }{\sqrt{2G+1}} S^{(5)}_{JGH} \ketbra{J, G+\tfrac12}{J, G+\tfrac12} \\\nonumber
    +&  \frac{\left|{\beta}\right|^2}{2(2G+1)}\,  S^{(6)}_{JGH} \ketbra{J, G+\tfrac12}{J, G+\tfrac12}
    \Bigg]  
    + h.c.
\end{align}
Recall that these operators are of the form $\ketbra{j^{\Sc\Gc\Hc}, j^{\Sc\Gc}}{j^{\Sc\Gc\Hc}, j'^{\Sc\Gc}} $. The coefficients $S^{(i)}_{JGH}$ are real numbers satisfying $-1 \leq S^{(i)}_{JGH} \leq 1$. They depend on the total angular momenta $j^{\Sc\Gc\Hc} = J$ and on the constant quantum numbers $j^\Gc = G$ and $j^\Hc = H$. The coefficients $S^{(i)}_{JGH}$ are defined in Appendix~\ref{app:A2}. 

Due to the use of the incoherent group average, we expect that the total angular momentum $J$ is in a probabilistic mixture. As we see in a moment, while the state \eqref{eq:twirled_rhoSGH} is clearly not pure, nevertheless, this relational description of the spin-1/2 $\mathcal{S}$ with respect to $\mathcal{G}$ and $\mathcal{H}$ has encoded information on the coherence of $\mathcal{S}$. We now turn to studying the scaling behavior of the CG coefficients $S^{(i)}_{JGH}$ in the regime of large and equal reference spins $G$ and $H$.

\subsection{Large reference spins limit}

We now take $G=H \gg 1$. The idea is that the usual state of a spin-1/2 should be recovered from the state \eqref{eq:twirled_rhoSGH}, since $\mathcal{G}$ and $\mathcal{H}$ would behave approximately as two classical gyroscopes pointing in orthogonal directions: the physical realization of two Cartesian axes. 

We verified this through a numerical investigation, which is given in detail in Appendix \ref{app:LargeLimit}. The main points are as follows. The coefficients  $S^{(i)}_{JGH}$ are all bounded from below by $-1$ and from above by $1$. Then, by inspection of \eqref{eq:twirled_rhoSGH}, when $G \gg 1$, the matrix elements corresponding to $S^{(4)}_{JGH}$, $S^{(5)}_{JGH}$ and $S^{(6)}_{JGH}$ will be negligible due to the pre-factors falling as $1/\sqrt{G}$, $1/\sqrt{G}$ and $1/G$ respectively. The matrix elements corresponding to $S^{(1)}_{JGH}$, $S^{(2)}_{JGH}$ and $S^{(3)}_{JGH}$ have pre-factors that fast become  constant when $G \gg 1$. Therefore, only these three terms are not negligible.

Then, the state $  \tilde{\mathcal{E}}^{C} (\rho^\mathcal{SGH})$ becomes approximately:
\begin{align}\label{eq:rho_SGH_lim}
    \tilde{\mathcal{E}}^{C} (\rho^\mathcal{SGH})
    &\approx \rho_{j^{\Sc\Gc\Hc}} \otimes \ket{\psi_{j^{\Sc\Gc}}}\bra{\psi_{j^{\Sc\Gc}}}
\end{align}
where
\begin{equation}
\label{eq:pureRelationalSpin}
    \ket{\psi_{j^{\Sc\Gc}}} = \alpha\ket{G + 1/2} + \beta\ket{G - 1/2}.
\end{equation}

This is our main result.  As expected, the density matrix $\rho_{j^{\Sc\Gc\Hc}}$, defined in Appendix~\ref{ap:calculation} and depicted in Fig.~\ref{fig:S1_JGG}, is diagonal in the total angular momentum $j^{\Sc\Gc\Hc}=J$. It is highly peaked on $J=\sqrt{2}G$, the classical magnitude when adding two orthogonal vectors of equal magnitude $G$. That is, as $G=H$ grows it approaches the pure state $\rho_{j^{\Sc\Gc\Hc}} \approx \ket{J=\sqrt{2}G} \bra{J=\sqrt{2}G} $. It holds classical information regarding the total angular momentum of the total system $\mathcal{SGH}$. The spin state also approaches the aligned/anti-aligned pure state \eqref{eq:pureRelationalSpin}, defined relationally with respect to the $j^{\Sc\Gc}=G\pm\frac12$ eigenstates. Analogously, had we followed the above procedure but first coupled $\mathcal{S}$ with $\mathcal{H}$, we would have arrived at a similar result to \eqref{eq:rho_SGH_lim}, with the spin-1/2 state being instead 
\begin{equation}
\label{eq:H}
\ket{\psi_{j^{\Sc\mathcal{H}}}} = \frac{\alpha+\beta}{\sqrt2} \ket{H + 1/2} + \frac{\alpha-\beta}{\sqrt2}\ket{H - 1/2}. 
\end{equation}

We have shown that the composite system $\mathcal{GH}$, used as a quantum reference system, suffices to fully reconstruct  the state of a qubit in the limit of large quantum numbers $G$ and $H$. This completes the calculation. Several comments are in order.

\begin{figure}
    \centering
    \resizebox{0.9\linewidth}{!}{\input{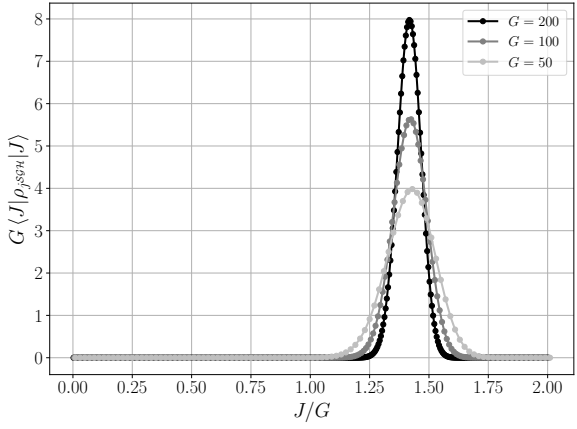}}
    \caption{Total angular momentum probability densities of the $\Sc\Gc\Hc$ system for $G=50,100,200$, as encoded by the density matrix $\rho_{j^{\Sc\Gc\Hc}}$ in \eqref{eq:rho_SGH_lim}. The peak occurs at $J\approx \sqrt{2}\,G$, consistent with the sum of two orthogonal vectors of equal lengths. Convergence to a delta function can be seen, indicating the classical limit for the reference system $\Gc\Hc$ (while $\Sc$ remains quantum).}
    \label{fig:S1_JGG}
\end{figure}

\section{Comments}\label{sec:comments}

\subsection{The classical limit}
\label{sec:clasLimit}

We saw in Section \ref{sec:one_directional reference} that when the reference system consists of only one large spin, the relational state for the spin-1/2 is a probabilistic mixture, see \eqref{eq:rhophys}. This makes it seem like an effective projective measurement was done, thereby `collapsing' the state \cite{Poulin_2006}. Moreover, Poulin \cite{Poulin_2006} arrived at a state of the form \eqref{eq:pureRelationalSpin} at an intermediate step of Sec.~II~B there, where the second large spin is a magnet $M$ driving the dynamics of the spin $\Sc$. However, this was not interpreted as a recovery of the state of $\Sc$ in a relational form, but as an example of the same `collapse' through $\Gc$ in presence of dynamics. Indeed, the magnet's size $M$ was taken much smaller than the gyroscope ($G=M^2\to\infty$) and then traced out, yielding a mixed state corresponding to the measurement of a precessing spin $\Sc$.
 
However, we are only manipulating the states through the (re)coupling theory of $SU(2)$, there is no interaction coupling: no measurement can be taking place. Instead, the quantum system used as a reference could not capture the information about the quantum coherence of the spin-1/2. For this, a second spin pointing in a  different direction is needed. In fact, while it was convenient for calculation purposes to take two spins that become orthogonal in the classical limit, a composite reference system of any two non-parallel large spins would also suffice (see Section \ref{sec:frameQuantisation}).

When two large spins are used as a composite reference system, the spin-1/2 can be described relationally with invariant (under rotations) quantum numbers. To extract information about $\Sc$, measurements can be done on the final density matrix. Concretely, a (tomographically) complete set of observables to describe $\Sc$ are the following Pauli-type spin operators:
\begin{align}
    \Sigma_1 &= \ketbra{G+1/2}{G-1/2} + \ketbra{G-1/2}{G+1/2} \label{eq:Sigma} \\\nonumber
    \Sigma_2 &= -i \ketbra{G+1/2}{G-1/2} + i\ketbra{G-1/2}{G+1/2} \\\nonumber
    \Sigma_3 &= \ketbra{G+1/2}{G+1/2} - \ketbra{G-1/2}{G-1/2}.
\end{align}
The observables $\Sigma_i$ are defined with respect to the joint angular momentum $j^\mathcal{SG} = G\pm1/2$ eigenstates of the subsystem $\mathcal{S}\mathcal{G}$, which make no reference to a system of axes. As before, a similar definition would hold with respect to $j^\mathcal{SH} = H\pm1/2$ if we had first coupled $\mathcal{H}$ to $\mathcal{S}$.

Let us now discuss the formal limit  $G=H \rightarrow \infty$. A technical point here is that for the limit of \eqref{eq:twirled_rhoSGH} to make sense, the bounded pre-factors $-1 \leq S^{(i)}_{JGH} \leq 1$ must converge. We give numerical evidence which strongly implies that this holds in Appendix \ref{app:LargeLimit}. Now, while the angular momenta $G$ and $H$ are infinite, the relational description remains meaningful: the Hilbert space is still $\mathbb{C}^2$. The two orthogonal states $\ket{G\pm1/2}$ then correspond to the states $\ket\uparrow$ and $\ket\downarrow$ of $\Sc$. 

Intuitively, the cone of uncertainty around the most probable direction that a reference spin is pointing towards shrinks to zero.  In particular, \eqref{eq:rho_SGH_lim} implies that the joint state becomes a pure state both for the fundamental spin and the composite reference system. For the latter, the state corresponds to a total angular momentum of $\sqrt2 G$, the length of the vector sum of the two orthogonal gyroscopes $\Gc$ and $\Hc$.  

In other words, although each gyroscope is treated fully quantum mechanically throughout, the limit $G,H\to\infty$ makes the spin states with distinct orientations effectively orthogonal, so that the reference systems have sharply distinguishable directions. Note that this holds in the limit whenever $\mathcal{G}$ and $\mathcal{H}$ point in directions differing by arbitrarily small angles. Therefore, as shown in Section \ref{sec:frameQuantisation}, we expect that any two non-parallel reference spins suffice to fully resolve the spin-1/2 state in the classical limit.

The above can be compared with the description of spin in a quantum reference frame considered in \cite{PhysRevResearch.3.043138}. In that work, a reference system was considered consisting of three orthogonal spins with infinite angular momentum from the outset, mimicking a Cartesian system of three axes. The reference system is not considered classical in that work in the sense that it can be imagined in a quantum superposition, but it does not encode any non-trivial commutation relations. Here, we have started from two spins with finite angular momentum, used recoupling theory, and showed that in the limit of infinite reference spins full information on the target spin is encoded.

In summary, in the classical limit the states $\ket{G\pm1/2}$ can be thought of abstractly as two states $\ket{A}$ and $\ket{AA}$, for aligned and anti-aligned, defining a qubit. The physical interpretation of the system in terms of relational quantities, is well defined and coincides with the usual description of a spin with respect to two classical orthogonal directions.

\subsection{Coherent vs incoherent average}

\label{sec:coherent}
Let us now examine how the calculation changes when using the coherent, rather than the incoherent, group average, and discuss the conceptual differences between the two approaches. We take $\Omega = \text{SU}(2)$, with normalized Haar measure $\text{d}\omega$. The coherent group average projects a vector $\ket{\psi}$ onto its invariant component:
\begin{equation}
    \ket{\psi} \mapsto \int_\Omega U(\omega) \ket{\psi} \;\text{d}\omega.
\end{equation} As an operation on density matrices, it reads
\begin{equation}
    \ketbra{\psi}{\psi} \mapsto\left(\int_\Omega U(\omega) \ket{\psi} \;\text{d}\omega \right) \left(\int_\Omega  \bra{\psi} U^\dag(\omega') \, \;\text{d}\omega'\right).
\end{equation}
This is different from the incoherent average that we have used, which corresponds to the diagonal group integration:
\begin{equation}
  \ketbra{\psi}{\psi} \mapsto \int_\Omega U(\omega) \ketbra{\psi}{\psi} U^\dag(\omega) \;\text{d}\omega.
\end{equation}

For the incoherent average, the sign change resulting from a $2 \pi$ rotation cancels when looking at the density operator, so the incoherent average can equivalently be expressed over $SO(3)$. In the case of the coherent average however, no such cancellation occurs, which is why we formulate the coherent average over $SU(2)$.

While the coherent average fixes the total angular momentum $j^{\Sc\Gc\Hc}$ to zero, the incoherent average does not in general fix $j^{\Sc\Gc\Hc}$. In this sense, the coherent average can be interpreted as treating the total angular momentum of the system as fiducial, a gauge effect which is to be done away with by projecting to the gauge invariant subspace.  Indeed, in order to attribute any physical significance to the total angular momentum of the system, it would need to be rotating with respect to something else. Therefore, in order to have a description that does not implicitly assume the existence of anything else other than the spin
and the two reference spins, the coherent average must be used. 

With the coherent average one arrives at a similar result, but for a different reference state. Projecting onto $j^{\Sc\Gc\Hc}=0$ requires $j^\Hc=j^{\Sc\Gc}=G\pm\tfrac12$: the angular momenta of $\Sc$, $\Gc$ and $\Hc$ must sum to zero, so $\Hc$ must be approximately as large as $\Gc$ and antiparallel to it. Similarly to the incoherent average, $\Gc$ still defines where `up' is and $\Hc$ where `right' is. However, the latter is now done through a coherent superposition of `down' and `up': the residual angular momentum of $\Gc\Hc$ points down for $j^\Hc=G+\tfrac12$ and up for $j^\Hc=G-\tfrac12$, and the relative phase between the two defines the positive $x$ direction. Namely, we take the state of $\Hc$ to be
\begin{align}\label{eq:chiH}
    \ket{\chi}^\Hc = \frac{1}{\sqrt2}\Big(\ket{G+\tfrac12,-G-\tfrac12}^\Hc_z + \ket{G-\tfrac12,-G+\tfrac12}^\Hc_z\Big),
\end{align}
so that the total state is $\left(\alpha \ket{\uparrow}^\Sc_z + \beta \ket{\downarrow}^\Sc_z\right) \otimes \ket{G,G}^\Gc_z \otimes \ket{\chi}^\Hc$. A relative phase $e^{-i\varphi}$ between the two terms of $\ket{\chi}^\Hc$ would correspond to rotating $\Hc$ by $\varphi$ about $z$. Only the first and third terms of \eqref{eq:SGxH} have a $j^{\Sc\Gc\Hc}=0$ component, and denoting with $P^0$ the projection on the invariant subspace,\footnote{The first and third terms of \eqref{eq:SGxH} pair with the $j^\Hc=G+\tfrac12$ and $j^\Hc=G-\tfrac12$ components of \eqref{eq:chiH}, respectively. For each we get the Clebsch--Gordan coefficient $\braket{0,0|j,m;j,-m}=(-1)^{j-m}/\sqrt{2j+1}$, i.e.\ $1/\sqrt{2G+2}$ and $1/\sqrt{2G}$, which times the coefficients in \eqref{eq:SGxH} yields \eqref{eq:P0chi}.}
\begin{align}\label{eq:P0chi}
    P^0\ket{\psi}^{\Sc\Gc\Hc} \propto \ket{j^{\Sc\Gc\Hc}=0}\otimes\Big(&\frac{\alpha}{\sqrt{2G+2}}\big|G+\tfrac12\big\rangle \nonumber\\
    &+ \frac{\beta}{\sqrt{2G+1}}\big|G-\tfrac12\big\rangle\Big),
\end{align}
which for large $G$ is proportional to
\begin{align}
\label{eq:cohRes}
    \ket{j^{\Sc\Gc\Hc} = 0} \otimes  \Big(
    \alpha \big|G+\tfrac{1}{2}\big\rangle
    +
    \beta \big|G-\tfrac{1}{2}\big\rangle
    \Big).
\end{align}
We can compare this to  \eqref{eq:rho_SGH_lim}, the result of the incoherent average with $\Hc$ in $\ket{H,H}^\Hc_x$. As we have seen, when $G = H \gg 1$, it reads approximately
\begin{equation}
\label{eq:incohRes}
     \ket{j^{\Sc\Gc\Hc} = \sqrt2 G} \otimes \Big(
    \alpha\,\big|G+\tfrac{1}{2}\big\rangle
    \;+\;
    \beta\,\big|G-\tfrac{1}{2}\big\rangle
    \Big)
 \end{equation}   

The reference state $\ket{H,H}^\Hc_x$ used for the incoherent average would not work here. For $H=G$ the state \eqref{eq:psiSGH} has no $j^{\Sc\Gc\Hc}=0$ component, and for a definite $H=G\pm\tfrac12$ only one of the two sectors survives.

To summarize, the incoherent group averaging diagonalizes over the total angular momentum of the entire system: it destroys coherence between the total angular momentum sectors of the entire system $\Gc\Hc\mathcal{S}$. This treats as classical the \emph{total} angular momentum of the system. However, as  \eqref{eq:rho_SGH_lim}, \eqref{eq:pureRelationalSpin}  and \eqref{eq:H} demonstrate, this does not imply that the coherence on \emph{subsystems} is erased: within a block of definite total angular momentum, we have coherent superpositions of the angular momenta states of the subsystems $\Gc\mathcal{S}$ and $\Hc\mathcal{S}$. As seen from \eqref{eq:cohRes} and \eqref{eq:incohRes}, using the coherent instead of the incoherent average has the effect that the total angular momentum is fixed to zero, instead of being in a probabilistic mixture that is highly peaked on the classically expected value, provided the second reference spin is replaced by one of uncertain magnitude, antiparallel to $\Gc$.

\subsection{Relation to reference frame quantization}
\label{sec:frameQuantisation}

We now discuss the relation of our construction to what is known as reference frame quantization. The idea here is as follows: given a symmetry group $\Omega$ with Haar measure $\text{d}\omega$, a target system $\Hc_S$ and a reference system $\Hc_R$ transforming under $U_S(\omega)$ and $U_R(\omega)$ respectively with $\omega \in \Omega$, the system $S$ is mapped onto the  $\Omega$-invariant subsystem of $\Hc_S \otimes \Hc_R$. More explicitly, the aim is to find maps from states $\rho_S$ and effects (observables) $E_S$
\begin{align}
    \rho_S &\mapsto \rho_{\inv},\\\nonumber
    E_S &\mapsto E_\inv, 
\end{align}
where $\rho_\inv = \mathcal{I}^\Omega(\rho_\inv)$ and $E_\inv = \mathcal{I}^\Omega(E_\inv)$ are $\Omega$-invariant operators on $\Hc_S \otimes \Hc_R$, and $\mathcal{I}^\Omega(\cdot) = \int_\Omega (U_S(\omega) \otimes U_R(\omega) ) (\cdot) (U_S^\dagger(\omega) \otimes U_R^\dagger(\omega) ) \;\text{d}\omega$. Moreover, we are interested in the notion of an \emph{exact encoding} of $\Hc_S$ in the $\Omega$-invariant subsystems of $\Hc_S \otimes \Hc_R$. This is the case when $\Tr(\rho_S E_S) = \Tr(\rho_\inv E_\inv)$, possibly in some limit.

A general procedure for implementing 
an exact encoding for compact groups $\Omega$ is proposed  in \cite{Bartlett_2007}. The reference systems are taken to be $\Hc_R \simeq L^2(\Omega)$ acted on by the left regular representation: $U_R(\chi)\ket \omega = \ket{\chi \omega}$ where $\braket{\omega|\chi} = \delta_{\omega,\chi}$ for $\omega,\chi \in \Omega$. The encoding is then:
\begin{align}
    \rho_S \mapsto  \rho_{\inv} &= \frac{1}{N} \int_\Omega U_S(\omega) \rho_S U_S^\dagger(\omega) \otimes \ketbra{\omega}{\omega}_R \;\text{d}\omega, \\\nonumber
    E_S \mapsto E_{\inv} &=  \int_\Omega U_S(\omega) E_S U_S^\dagger(\omega) \otimes \ketbra{\omega}{\omega}_R \;\text{d}\omega,
\end{align}
where $N$ is some normalisation constant $N \I= \int \ketbra{g}{g}\ \text{d}g$ (for a  rigorous treatment of this integral see e.g.~\cite{Loveridge_2018}). 

 Taking $\Omega = \mathrm{SU}(2)$, the procedure sketched above is different from the one that we have presented, as we have made use of reference frames that are not isomorphic to $L^2(\SU(2))$. A precise group representation theoretic characterisation of the $\Gc
\Hc$ reference system used in this work, and how it contrasts to $L^2(\SU(2))$ reference frames, is given in Appendix~\ref{app:CSS_characterisation}. 

We find it difficult to imagine how to physically model the quantum reference system as living in $L^2(\SU(2))$.  Considering instead a composite reference system made of two quantum spins, as we have done in this work, the interpretation of the relational state becomes straightforward. Furthermore, the description is quantitative, the formulas we have arrived at are explicit and could be used to calculate numerically e.g.~the amount of coherence in the spin-1/2 sector for given $G$ and $H$. 

On the other hand, using techniques inspired from the formal apparatus developed in \cite{Bartlett_2007}, we can make an abstract but general statement regarding the case of a system of interest with many degrees of freedom. We have established here that a pair of gyroscopes $\Gc,\Hc$ can fully resolve a single spin-1/2 state in the classical limit $G,H \to \infty$. We now ask, does this result extend to qudits, which carry spin $\frac{d-1}{2}$ representations of $\SU(2)$? Indeed, we show in Appendix~\ref{app:coherent_state_encoding} that for any compact group $\Omega$ and coherent state system $\{\ket{\psi(\omega)}|\omega \in \Omega\}$ where $\ket{\psi(\omega)} \not\sim \ket{\psi(\chi)}$ for $\chi \neq \omega$, if the weight $\mu$ defined there tends to a delta function at the identity, in this limit we have an exact encoding: $\Tr(\rho_{\inv} E_\inv) \to \Tr(\rho_S E_S)$. In the language of Ref.~\cite{Loveridge_2018}, $\Tr(\rho_\inv E_\inv)$ is the expectation value in $\rho_S$ of the relativised effect restricted to the reference state, and the classical limit is the high-localisation limit of the reference state --- see Appendix~\ref{app:coherent_state_encoding} for details. This holds true for the type of states we have used to model the reference system $\mathcal{GH}$, of the form $U(\omega)\big(\ket{G,G}^\Gc_z\otimes\ket{H,H}^\Hc_x\big)$. They can be chosen to form an overcomplete coherent state basis and so that for any two different $\omega$ they become orthogonal in the limit of large quantum numbers $G$ and $H$, see e.g.~\cite{perelomov_generalized_1986} for details. 

Therefore, the general moral of the calculation carries through to target systems of many degrees of freedom: a reference quantum system with two non-commuting operators satisfying the algebra of angular momenta increasingly resolves the quantum state of a target system that transforms under $SU(2)$ as the quantum numbers of the reference grow larger, and does so exactly in the limit.

\section{Conclusion}

We showed how a spin-$1/2$ can be fully described in relation to another quantum system. The point of departure was to write in the standard quantum mechanical formalism the joint state of a qubit $\mathcal{S}$ with two other quantum mechanical spins $\mathcal{G}$ and $\mathcal{H}$ that are pointing in orthogonal directions and have arbitrary angular momenta. We calculated the state after the incoherent group average over $\SO(3)$, which removes any reference to external fixed axes. The directional information that remains is relational, encoding quantum amplitudes on the degree that $\mathcal{S}$ is aligned or anti-aligned with the reference angular momenta. We demonstrated that when the quantum numbers of the reference systems are large, the usual state of the qubit is recovered, thereby demonstrating that the encoding becomes exact as the classical limit for the reference is approached. 

We clarified in this context some points of contention in the contemporary discussion on quantum reference frames regarding when it is appropriate to use the incoherent or the coherent average. The incoherent group averaging does \emph{not} result in loss of coherence of the target system---here the spin-1/2. It is the coherence between sectors of different \emph{total} angular momentum that is lost; the coherence between the angular momenta of subsystems is encoded. Only the latter information is what is relevant if what is sought after is to remove any reference to a background. The incoherent average retains information about the coherence of subsystems, but yields a non-zero total angular momentum for the entire system. This implies that a background system is needed, in order to physically interpret the total rotation of the joint system.

 Whether to use the incoherent versus the coherent average depends on whether the total angular momentum of the system is to be treated as physical or not. For a completely background independent relational description, the coherent average should be used. This also would be appropriate, for instance, if the operators satisfying the algebra of angular momenta do not physically correspond to rotations but to areas, as happens in the theory of spin-networks forming the kinematics of loop quantum gravity. Indeed, in that theory, the coherent average is used (in conjunction with a closure constraint) to define quantum polyhedral geometries, see e.g.~\cite{Rovelli:2014ssa,Bianchi:2010gc}. The reference systems in this context could be normals to the areas determined by the quantized geometry, and the components of a spin relative to them were used in \cite{Sahlmann_2025} to construct Bell-violating states in loop quantum gravity. For large reference spins, the spin component relative to a normal acts like $\Sigma_3$ in \eqref{eq:Sigma}.

 We have seen that the coherent average also recovers the qubit in the limit of large quantum numbers, but with a different reference: instead of two gyroscopes along orthogonal axes, the second must be antiparallel to the first, in a coherent superposition of magnitudes $G\pm\frac12$ whose relative phase fixes the transverse direction. Beyond this, the coherent average forces a zero total angular momentum while the incoherent average fixes it to the classically expected value. 
 
Finally, we have shown that the general conclusion of our case study for a spin-1/2 target system, carries through for arbitrarily large target systems that transform under $SU(2)$: a quantum reference system with two operators that do not commute and satisfy the algebra of angular momenta, will yield an exact encoding of the target system in the limit of large quantum numbers for the reference. Regarding future steps, it is of particular interest to  study how well the encoding works for more general reference states (entangled, mixed or non-coherent spin states), thereby quantifying how imperfections in the quantum reference frame degrade the recovered coherence.

\bibliography{refs}

\onecolumngrid
\clearpage
\appendix
\section{Background independent spin given two directional references}\label{ap:calculation}

\subsection{Basis Change }
\label{app:A1}
To write the state of the system entirely in the $z$-basis we use the basis change \eqref{eq:Hx=Hz}, given by
\begin{align}
\ket{H, H}^{\Hc}_x  = \frac{1}{2^H} \sum_{h\,=-H}^{H} \sqrt{\binom{2H}{H+h}}  \ket{H, h}^{\Hc}_z. \nonumber
\end{align}
Replacing this in the state \eqref{eq:SGxH}, which is given by
\begin{align}
\ket{\psi}^{\Sc\Gc\Hc} = \Big(\alpha \ket{G+\tfrac{1}{2}, G+\tfrac{1}{2}} + \frac{\beta}{\sqrt{2G+1}} \ket{G+\tfrac{1}{2}, G-\tfrac{1}{2} } + \frac{\beta\,\sqrt{2G}}{\sqrt{2G+1}} \ket{G-\tfrac{1}{2}, G-\tfrac{1}{2}} \Big)\otimes \frac{1}{2^H} \sum_{h\,=-H}^{H} \sqrt{\binom{2H}{H+h}}  \ket{H, h}^{\Hc}_z \nonumber
\end{align}
we obtain
\begin{align}
\ket{\psi}^{\Sc\Gc\Hc} = \frac{1}{2^H} \sum_{h\,=-H}^{H} \sqrt{\binom{2H}{H+h}} \Big(\alpha \ket{G+\tfrac{1}{2}, G+\tfrac{1}{2}} + \frac{\beta}{\sqrt{2G+1}} \ket{G+\tfrac{1}{2}, G-\tfrac{1}{2} } + \frac{\beta\,\sqrt{2G}}{\sqrt{2G+1}} \ket{G-\tfrac{1}{2}, G-\tfrac{1}{2}} \Big)\otimes \ket{H, h}^{\Hc}_z.
\end{align}

Next, we do the basis change $\left\{\ket{j^{\Sc\Gc}, m^{\Sc\Gc}}\right\} \otimes \left\{\ket{ j^\Hc,m^\Hc}\right\} \rightarrow   \left\{\ket{j^{\Sc\Gc\Hc}, m^{\Sc\Gc\Hc}, j^{\Sc\Gc}}\right\}$ with the Clebsch-Gordan coefficients defined in \eqref{eq:cg}. This yields the following three terms.

\noindent First term:
\begin{align}
    \sum_{h\,=-H}^{H} \alpha  \sqrt{\binom{2H}{H+h}}  \ket{G+\tfrac{1}{2}, G+\tfrac{1}{2} }\ket{H, \,h}
    = \sum_{h\,=-H}^{H} \alpha  \sqrt{\binom{2H}{H+h}}
    \sum_{J=G+1/2 + h}^{G + 1/2 + H} C^{J, G+1/2 + h }_{G+1/2, G+1/2, h} \ket{J, G+\tfrac12 + h , G + \tfrac12}.
\end{align}
Second term:
\begin{align}
    &\sum_{h\,=-H}^{H} \frac{\beta}{\sqrt{2G+1}}  \sqrt{\binom{2H}{H+h}}  \ket{G+\tfrac{1}{2}, G-\tfrac{1}{2} }\ket{H, \,h}\\\nonumber
    =&\sum_{h\,=-H}^{H} \frac{\beta}{\sqrt{2G+1}}  \sqrt{\binom{2H}{H+h}}
    \sum_{J=G-1/2 + h}^{G + 1/2 + H} C^{J, G-1/2 + h }_{G+1/2, G-1/2, h} \ket{J, G-\tfrac12 + h , G + \tfrac12},
\end{align}
Third term:
\begin{align}
    &\sum_{h\,=-H}^{H} \frac{\beta \sqrt{2G}}{\sqrt{2G+1}}  \sqrt{\binom{2H}{H+h}}  \ket{G-\tfrac{1}{2}, G-\tfrac{1}{2} }\ket{H, \,h} \\\nonumber
    = &\sum_{h\,=-H}^{H} \frac{\beta\sqrt{2G}}{\sqrt{2G+1}}  \sqrt{\binom{2H}{H+h}}
    \sum_{J=G-1/2 + h}^{G - 1/2 + H} C^{J, G-1/2 + h }_{G-1/2, G-1/2, h} \ket{J, G-\tfrac12 + h , G - \tfrac12}.
\end{align}
Their sum, including the $1/2^{H}$ factor, is the total state as given in \eqref{eq:psiSGH}:
\begin{align}
    \ket{\psi}^{\Sc\Gc\Hc} =\frac{1}{2^H} \sum_{h=-H}^{H} \sum_{J=G-1/2+h}^{G+1/2+H} \sqrt{\binom{2H}{H+h}}
    \Bigg[ &\alpha \, C^{J, G+1/2+h}_{G+1/2,  G+1/2 , h} \ket{J, G+\tfrac{1}{2}+h, G+\tfrac{1}{2}} \\\nonumber
    + \, &\frac{\beta}{\sqrt{2G+1}}  C^{J, G-1/2+h}_{G+1/2, G-1/2,  h} \ket{J, G-\tfrac{1}{2}+h, G+\tfrac{1}{2}} \\\nonumber
    + \, &\frac{\sqrt{2G} \, \beta}{\sqrt{2G+1}}  C^{J, G-1/2+h}_{G-1/2, G-1/2 , h} \ket{J, G-\tfrac{1}{2}+h, G-\tfrac{1}{2}} \Bigg]. \nonumber
\end{align}

\subsection{Group Averaging}
\label{app:A2}
Now, we will apply the incoherent group average to the state's density matrix $\rho^\mathcal{SGH}$ corresponding to the pure state $\ket{\psi}^{\Sc\Gc\Hc}$ given in \eqref{eq:psiSGH}. Explicitly, $\rho^\mathcal{SGH}$ is given by
\begin{align}\label{eq:density_matrix_SGH}
    \rho^\mathcal{SGH} = \frac{1}{2^H} \sum_{h=-H}^{H} \sum_{J=G-1/2+h}^{G+1/2+H}
    \sqrt{\binom{2H}{H+h}} \Bigg[ &\alpha \; C^{J, G+1/2+h}_{G+1/2, G+1/2, h} \ket{J, G+1/2+h, G+1/2} \\\nonumber
    + \, &\frac{\beta}{\sqrt{2G+1}}  C^{J, G-1/2+h}_{G+1/2, G-1/2, h} \ket{J, G-1/2+h, G+1/2} \\\nonumber
    + \, &\frac{\sqrt{2G} \, \beta}{\sqrt{2G+1}}  C^{J, G-1/2+h}_{G-1/2, G-1/2, h} \ket{J, G-1/2+h, G-1/2} \Bigg] 
    \\\nonumber
    \frac{1}{2^H} \sum_{h'=-H}^{H} \sum_{J'=G-1/2+h'}^{G+1/2+H}
    \sqrt{\binom{2H}{H+h'}} \Bigg[ &\bar\alpha \; \bar{C}^{J', G+1/2+h'}_{G+1/2, G+1/2, h'} \bra{J', G+1/2+h', G+1/2} \\\nonumber
    + \, &\frac{\bar\beta}{\sqrt{2G+1}}  \bar{C}^{J', G-1/2+h'}_{G+1/2, G-1/2, h'} \bra{J', G-1/2+h', G+1/2} \\\nonumber
    + \, &\frac{\sqrt{2G} \, \bar\beta}{\sqrt{2G+1}}  \bar{C}^{J', G-1/2+h'}_{G-1/2, G-1/2, h'} \bra{J', G-1/2+h', G-1/2} \Bigg].
\end{align}
Due to Schur's Lemma, when $J_1 \neq J_2$ or $m_1 \neq m_2$ the matrix elements $\ketbra{J_1, m_1}{J_2, m_2}$ vanish. The remaining, non-zero matrix elements, are those for which $J_1 = J_2$ and $m_1 = m_2$. When calculating $\mathcal{E}(\rho^{\Sc\Gc\Hc}) =  \int_{SO(3)} \text{d}\mu(\Omega) \;R^{\Sc\Gc\Hc}(\Omega)\; \rho^{\Sc\Gc\Hc} \; R^{\Sc\Gc\Hc}(\Omega)^\dag$, this means that one has to match $J=J'$ for every element depending on the specific entry choose $h' = h$ or $h' = h \pm 1$. Performing the average then leads to \eqref{eq:twirled_rhoSGH}
\begin{align}
    \mathcal{E}(\rho^\mathcal{SGH}) =\sum_{J = G-1/2-H}^{G+1/2+H}  \Bigg[  
    &\frac{\left|\alpha\right|^2}{2}\,  S^{(1)}_{JGH} \ketbra{J, G+\tfrac12}{J, G+\tfrac12} 
    +  \frac{\sqrt{2G}\alpha \Bar{\beta}}{\sqrt{2G+1}} S^{(2)}_{JGH} \ketbra{J, G+\tfrac12}{J, G-\tfrac12}  
    \nonumber \\\nonumber
    +  &\frac{G \, \left|{\beta}\right|^2}{2G+1}\,  S^{(3)}_{JGH} \ketbra{J, G-\tfrac12}{J, G-\tfrac12} 
    +  \frac{\sqrt{2G} \, \left|{\beta}\right|^2}{2G+1} \, S^{(4)}_{JGH} \ketbra{J, G+\tfrac12}{J, G-\tfrac12}
    \\\nonumber
    + & \frac{\alpha \Bar{\beta} }{\sqrt{2G+1}} S^{(5)}_{JGH} \ketbra{J, G+\tfrac12}{J, G+\tfrac12} 
    +  \frac{\left|{\beta}\right|^2}{2(2G+1)}\,  S^{(6)}_{JGH} \ketbra{J, G+\tfrac12}{J, G+\tfrac12}
    \Bigg]  
    + h.c., \nonumber
\end{align}
with
\begin{align}
 S^{(1)}_{JGH}= & \frac{1}{4^H}  \sum_{h= -H}^{H} \binom{2H}{H+h} |C^{J, G+1/2+h}_{G+1/2, G+1/2, h}|^2  \\\nonumber
 S^{(2)}_{JGH}=  & \frac{1}{4^H}  \sum_{h= -H}^{H}\sqrt{\binom{2H}{H+h}\binom{2H}{H+h+1}} C^{J, G+1/2+h}_{G+1/2, G+1/2, h} \bar{C}^{J,G+1/2+h}_{G-1/2, G-1/2, h+1}
  \\\nonumber
   S^{(3)}_{JGH}=  & \frac{1}{4^H}  \sum_{h= -H}^{H}\binom{2H}{H+h} |C^{J, G-1/2+h}_{G-1/2, G-1/2, h}|^2 \\\nonumber
   S^{(4)}_{JGH}=  & \frac{1}{4^H}  \sum_{h= -H}^{H}\binom{2H}{H+h}\bar{C}^{J, G-1/2+h}_{G-1/2, \, G-1/2, h} C^{J, G-1/2+h}_{G+1/2, G-1/2, h}\\\nonumber
   S^{(5)}_{JGH}=  & \frac{1}{4^H}  \sum_{h= -H}^{H}\sqrt{\binom{2H}{H+h}\binom{2H}{H+h+1}} C^{J, G+1/2+h}_{G+1/2, G+1/2, h} \bar{C}^{J, G+1/2+h}_{G+1/2, G-1/2, h+1}\\\nonumber
   S^{(6)}_{JGH}=  & \frac{1}{4^H}  \sum_{h= -H}^{H}\binom{2H}{H+h}|C^{J, G-1/2+h}_{G+1/2, G-1/2, h}|^2. 
\end{align}

\subsection{Large Gyroscope Limit}
\label{app:LargeLimit}
Note that even though the Clebsch-Gordan coefficients are all real numbers, we keep the complex conjugate notation for bookkeeping purposes. The Clebsch-Gordan coefficients are real numbers between -1 and 1 since they are inner products of normalized vectors. This implies that the sums $S^{(i)}_{JGH}$ defined above are also bounded by -1 from below and 1 from above:
\begin{align}
\left| S^{(i)}_{JGH} \right|  \leq \frac{1}{4^H}  \sum_{h= -H}^{H}   \binom{2H}{H+h} = \frac{1}{4^H} \cdot 2^{2H} = 1.
\end{align}
Note that for $i=2,5$, the binomial coefficient is replaced with $\sqrt{\binom{2H}{H+h}\binom{2H}{H+h+1}}$, but this does not influence the limit, as can be confirmed numerically.

When $G=H\to\infty$, the matrix elements in $\mathcal{E}(\rho^\mathcal{SGH})$ are of the form $a^{(i)}_G S^{(i)}_{JGH}$. For $i=4,5,6$, we have that $a^{(i)}_G \to 0$. Since $S^{(i)}_{JGH}$ are bounded, the corresponding matrix element vanishes, that is, $a^{(4,5,6)}_G S^{(4,5,6)}_{JGH} \to 0$. 

\begin{figure}
    \centering
    \resizebox{0.6\linewidth}{!}{\input{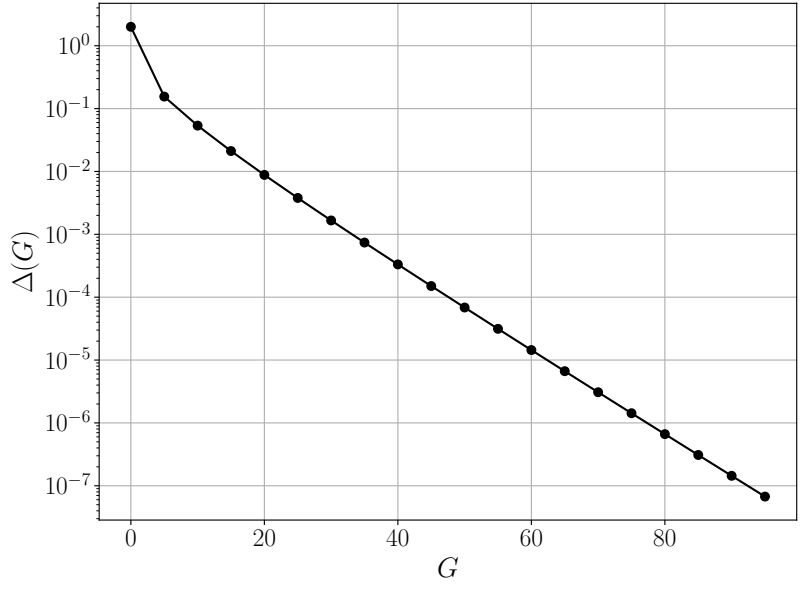}}
    \caption{Numerical evaluation of $\Delta(G) = \sum_J (|S^{(1)}_{JGG}-S^{(2)}_{JGG}| + |S^{(1)}_{JGG}-S^{(3)}_{JGG}|)$ as a function of $G$. The decay of this value with increasing $G$ provides numerical evidence that $S^{(1)}_{JGG}$, $S^{(2)}_{JGG}$, and $S^{(3)}_{JGG}$ converge to a common asymptotic limit, denoted $S_{J\infty}$.}
    \label{fig:diff}
\end{figure}

For the remaining matrix elements $i=1,2,3$, we denote $ S^{(1,2,3)}_{JGH} \to S^{(1,2,3)}_{J\infty}$. We work under the assumption that this limit exists. We give numerical evidence that this is the case at the end of this Appendix. The incoherently group averaged state in the limit $G=H \to \infty$ is therefore given by
\begin{align*}
    \mathcal{E}(\rho^\mathcal{SGH}) \to  \mathcal{E}^{\lim} (\rho^\mathcal{SGH}) =  
    \sum_{J = 1/2}^\infty  &\Bigg[  
    \frac{|\alpha|^2}{2}\,  S^{(1)}_{J \infty} \ketbra{J, G+1/2}{J, G+1/2} \\
    &+ \frac{|\beta|^2}{2}\,  S^{(3)}_{J \infty} \ketbra{J, G-1/2}{J, G-1/2} \\
    &+ \alpha\bar\beta \, S^{(2)}_{J \infty} \ketbra{J, G+1/2}{J, G-1/2} 
    \Bigg]  + h.c.
    \\
\end{align*}
The letter $G$ remains here not as a variable, but only as part of the conventional label for the two different eigenstates of the operator $j^{\mathcal{S}\mathcal{G}}$. Furthermore, looking at the expressions for $S^{(i)}_{JGH}$ one notices the numerically verifiable fact that for all $J$, $S^{(1)}_{J\infty}=S^{(2)}_{J\infty}=S^{(3)}_{J\infty}$ (see Fig.~\ref{fig:diff}), which we denote as $S_{J\infty}$. We can therefore further compactify the expression for the limit physical state to the form \eqref{eq:rho_SGH_lim},
\begin{align}
    \mathcal{E}^{\lim} (\rho^\mathcal{SGH}) &= 
    \rho_{j^{\Sc\Gc\Hc}} \otimes \rho_{j^{\Sc\Gc}} 
    =\rho_{j^{\Sc\Gc\Hc}} \otimes \ket{\psi_{j^{\Sc\Gc}}}\bra{\psi_{j^{\Sc\Gc}}}, \\\nonumber
    \ket{\psi_{j^{\Sc\Gc}}} &= \alpha\ket{G + 1/2} + \beta\ket{G - 1/2}.
\end{align}
with the constant, numeric density matrix
\begin{align}\label{eq:rhoJ}
\rho_{j^{\Sc\Gc\Hc}} = \sum_{J = 1/2}^\infty  S_{J \infty}\ketbra{J}{J},
\end{align}
describing the total angular momentum of the entire system $\Sc\Gc\Hc$. Its values $ \bra J \rho_{j^{\Sc\Gc\Hc}} \ket J=  S_{J \infty}$ are depicted in Fig.~\ref{fig:S1_JGG} and show convergence to a delta function peaked at $J = \sqrt2\, G$ as $G\to\infty$ corresponding to the size of the sum of the two orthogonal spins $\Gc$ and $\Hc$. This constitutes numerical evidence for the existence of the limit $S_{J \infty}$.

\section{Coherent state characterisation of quantum reference frames}

In this appendix we contrast the reference frame constructed out of gyroscopes used in the present work to the $L^2(\SU(2))$ reference frame used in the reference frame quantization procedure of~\cite{Bartlett_2007}. We show that these are two different group coherent state systems~\cite{perelomov_generalized_1986}, even though both allow for a full encoding of a qubit in the relative degrees of freedom between the reference frame and spin 1/2 particle.  The $L^2(G)$ reference frame is ideal, in the sense that it allows for a perfect encoding of the system of interest in the $G$-invariant algebra of the composite of reference and system. Examples of reference frame quantization in the literature which explore the relation between the size of the reference frame and the accuracy of the encoding, as done in the present work, include~\cite{Miyadera2016,Skotiniotis2017,Loveridge2020}.

\subsection{Ideal \texorpdfstring{$L^2(\SU(2))$}{L2(SU(2))} coherent state systems}

For a compact group $G$ the space of square integrable functions $L^2(G)$ (for a given choice of measure) transforms under the left regular representation as:
\begin{align}
    U_L(g)\ket{g'} = \ket{gg'}.
\end{align}

Using the Peter-Weyl theorem we can decompose $L^2(G)$ under the left regular representation of $G$ as:
\begin{align}
    L^2(G) \simeq \bigoplus_{\lambda \in \hat G} V_\lambda \otimes W_\lambda, 
\end{align}
where $\hat G$ is the set of irreducible representations of $G$, $V_\lambda$ carries the irreducible representation $U_\lambda$ of $G$, and the \emph{multiplicity} space $W_\lambda$ carries $\dim(W_\lambda)$ copies of the trivial representation. Moreover, $\dim(V_\lambda) = \dim(W_\lambda)$. In the case of $G = \SU(2)$ the irreps are labeled by half-integers $\lambda$ and hence parameterizing a rotation $\Omega \in \SU(2)$ by three real parameters $\alpha, \beta, \gamma$ gives:
\begin{align}
    U_L(\alpha, \beta, \gamma) \simeq \bigoplus_{j \in \frac12\Nl} D^j(\alpha,\beta, \gamma) \otimes \I_{2 j + 1}.
\end{align}
The states $\ket{g}$ have support in every $j \in \frac12\Nl$ and hence the coherent state system transforms under $\bigoplus_{j \in \frac12\Nl} D^j(\alpha,\beta, \gamma) \otimes \I_{2 j + 1}$.

As shall be proven in the next section a key difference with the $\SO(3)$ coherent state system constructed in the present work is that the $L^2(\SU(2))$ coherent state system also carries a unitary representation of the right regular action, which in this case is the right regular representation:
\begin{align}
    U_R(g) \ket{g'} = \ket{g' g^{-1}},
\end{align}
where once more using the Peter-Weyl theorem:
\begin{align}
    U_R(\alpha, \beta, \gamma) \simeq \bigoplus_{j \in \frac12\Nl}   \I_{2 j + 1} \otimes \bar D^j(\alpha,\beta, \gamma).
\end{align}

\subsection{Orthogonal gyroscope coherent state systems}\label{app:CSS_characterisation}

\subsubsection{Representation theoretic characterisation}

The quantum reference frame of Sec.~\ref{sec:two_directional references} is obtained from the reference state $ \ket{\psi(I)}^{\Gc\Hc}  = \ket{G,G}^\Gc_z \otimes \ket{H,H}^\Hc_x$ acted on by the representation $D^G(\Omega) \otimes D^H(\Omega)$:  $\ket{\psi(\Omega)}^{\Gc\Hc} = D^G(\Omega) \otimes D^H(\Omega)  \ket{\psi(I)}^{\Gc\Hc}  $, $\Omega \in \SO(3)$. 

Using $\SU(2)$ representation theory we can decompose the representation as follows:
\begin{align}
        V^G \otimes V^H  &\simeq \bigoplus_{j = |G - H|}^{G + H} V^j, \label{eq:irrep_decomp_GH}\\\nonumber
       D^G(\Omega) \otimes D^H(\Omega) &\simeq \bigoplus_{j = |G - H|}^{G + H} D^j(\Omega) .
\end{align}
We now determine the support of the reference state $\ket{G,G}^\Gc_z \otimes \ket{H,H}^\Hc_x$ on the subspaces $V^j$ carrying representations $D^j$.
Using standard angular momentum coupling rules we have:
\begin{align}
    \ket{\psi(I)}^{\Gc\Hc} =  \ket{G,G}^\Gc_z \otimes \ket{H,H}^\Hc_x &=\ket{G,G}^\Gc_z \otimes \left( \frac{1}{2^H} \sum_{h\,=-H}^{H} \sqrt{\binom{2H}{H+h}}  \ket{H, h}^{\Hc}_z\right) \\ \nonumber
     &\simeq \frac{1}{2^H} \sum_{h=-H}^H \sqrt{\binom{2H}{H+h}} \sum_{j=|G-H|}^{G+H} C^{j, G+h}_{G, G; H, h} \ket{j, G+h}\\\nonumber
     &=  \frac{1}{2^H}  \sum_{j=|G-H|}^{G+H}\sum_{h=-H}^H \sqrt{\binom{2H}{H+h}} C^{j, G+h}_{G, G; H, h} \ket{j, G+h}.
\end{align}

For every $j \in \{|G-H|,..., G+H\}$ one has a $h \in \{-H,...,H \}$ such that $j = G+ h$, hence for every $j$ there is a coefficient $C^{j, G+h}_{G, G; H, h} = C^{j, j}_{G, G; H, h} \neq 0$. This shows that $\ket{\psi(I)}^{\Gc\Hc}$ has support in every irrep $D^j$.

Since $\ket{\psi(I)}^{\Gc\Hc} $ has support in every irreducible representation $D^j(R)$ for $j \in \{|G-H|,..., G+H\}$ the coherent state system $\{\ket{\psi(\Omega)}^{\Gc\Hc}|\Omega \in \SO(3)\}$ transforms under the representation $D^G(\Omega) \otimes D^H(\Omega) \simeq \bigoplus_{j = |G - H|}^{G+H} D^j(\Omega)$ of $\SO(3)$.

To confirm that it is a $\SO(3)$ coherent state system we need to show that:

\begin{align}
    \ket{\psi(\Omega)} = e^{i \phi} \ket{\psi(\Omega')} \implies \Omega = \Omega', \forall \Omega,\Omega' \in \SO(3).
\end{align}

Equivalently we need to show that the stabilizer group of the ray $\ket{\psi(I)}$ is trivial. It is sufficient to show that the stabilizer group of the vector  $\ket{\psi(I)}$ is trivial.

The stabilizer of $\ket{G,G}^\Gc_z $ is the ${\rm U}(1)$ subgroup ${\rm Stab}(\ket{G,G}^\Gc_z ) = D^G(e^{i Zt}), t \in [0,2\pi)$ and the  stabilizer of $\ket{H,H}^\Hc_x $ is the ${\rm U(1)}$ subgroup ${\rm Stab}(\ket{H,H}^\Hc_x) =D^H(e^{i Xt}), t \in [0,2\pi)$. The stabilizer group ${\rm Stab}(\ket{G,G}^\Gc_z \otimes \ket{H,H}^\Hc_x)$ of $\ket{\psi(I)}^{\Gc\Hc} =  \ket{G,G}^\Gc_z \otimes \ket{H,H}^\Hc_x$ ,  is given by ${\rm Stab}(\ket{G,G}^\Gc_z ) \cap {\rm Stab}(\ket{H,H}^\Hc_x) = D^G(I) \otimes D^H(I)$.

\subsubsection{Left-right regular action}

The $\SO(3)$ coherent state system $\ket{\psi(\Omega)}$ carries a left regular action (which is not the left regular representation however):
\begin{align}
    D^G(\Omega) \otimes D^H(\Omega) \ket{\psi(\Omega')} = \ket{\psi(\Omega\Omega')}, \forall \Omega, \Omega' \in \SO(3).
\end{align}

We now show that the right regular action:
\begin{align}\label{eq:rightaction}
    \ket{\psi(\Omega')} \mapsto \ket{\psi(\Omega'\Omega^{-1})},
\end{align}
is not unitary.

Let us assume there exists a unitary representation $C: \Omega \mapsto C(\Omega)$ such that $C(\Omega)\ket{\psi(\Omega')} = \ket{\psi(\Omega' \Omega^{-1})}$. We now show that this leads to a contradiction.

By construction this representation commutes with $D(\Omega) = D^G(\Omega) \otimes D^H(\Omega)$. By Schur's lemma this implies that:
\begin{align}
    C(\Omega) \simeq \bigoplus_{j = |G - H|}^{G + H} c_j(\Omega) \I^j, c_j(\Omega) \in \Cl \ \forall j \in \{|G-H|,..., G+H\}. 
\end{align}
By unitarity $c_j(\Omega) = e^{i \theta_j(\Omega)}$ for all $ \forall j \in \{|G-H,..., G+H\}, \Omega \in \SO(3)$.

 Eq.~\eqref{eq:rightaction} implies that $C(\Omega)\ket{\psi(I)} = \ket{\psi(\Omega^{-1})}$. Define $\Pi_j$ the projector onto $V_j$, the carrier space of the irreducible representation $D_j$. Then $\Pi_j(C(\Omega)\ket{\psi(I)}) = c_j(\Omega) \Pi_j(\ket{\psi(I)})$. However by assumption $D(\Omega^{-1}) \ket{\psi(I)} = C(\Omega)\ket{\psi(I)}$ which implies $\Pi_j(D(\Omega^{-1})\ket{\psi(I)}) = D^j(\Omega^{-1}) \Pi_j \ket{\psi(I)} =  c_j(\Omega) \Pi_j(\ket{\psi(I)})$. This implies the one dimensional subspace spanned by $\Pi_j(\ket{\psi(I)})$ is invariant under $D^j(\Omega)$ which contradicts the fact that $D^j$ is irreducible and of dimension strictly greater than 1 for $j \neq 0$.

\subsection{Reference frame quantization for generic coherent state system}\label{app:coherent_state_encoding}

We now consider a reference frame $\Hc_R$ spanned by some coherent state system for $G$: $\{\ket{\psi(g)}|g \in G\}$ which is not ideal: $|\!\braket{\psi(g)|\psi(g')}\!|^2\neq 0$ for $g \neq g'$. Take the normalized Haar measure $\int_G \text{d}g = 1$. The operator
\begin{align}\label{eq:overcomplete}
    F := \int_G \ketbra{\psi(g)}{\psi(g)} \;\text{d}g
\end{align}
is then positive and invertible, and commutes with every $U_R(g)$. It is in general not proportional to the identity: for the gyroscopes of Appendix~\ref{app:CSS_characterisation}, Schur's lemma gives $F = \sum_j \frac{p_j}{2j+1}\Pi_j$ with $p_j = \bra{\psi(I)}\Pi_j\ket{\psi(I)}$, i.e.\ $F$ rescales every $j$ sector. We therefore use the states $\ket{\tilde\psi(g)} = F^{-1/2}\ket{\psi(g)}$ instead. That is, we use the following covariant POVM to define the frame's orientation
\begin{align}
    \gamma_R(g) := \ketbra{\tilde\psi(g)}{\tilde\psi(g)} = F^{-1/2}\ketbra{\psi(g)}{\psi(g)}F^{-1/2}, \qquad U_R(h)\,\gamma_R(g)\,U_R^\dagger(h) = \gamma_R(hg), \qquad \int_G \gamma_R(g)\;\text{d}g = \I .
\end{align}
When $F = N\I$ it reduces to $\gamma_R(g) = \frac1N \ketbra{\psi(g)}{\psi(g)}$.

The encoding maps are:
\begin{align}
      \rho_S \mapsto \rho_{\inv} :=& \int_G U_S(g) \rho_S U_S^\dagger(g) \otimes \ketbra{\psi(g)}{\psi(g)}_R \;\text{d}g, \\\nonumber
    E_S \mapsto E_{\inv} :=& \int_G U_S(g) E_S U_S^\dagger(g) \otimes \gamma_R(g) \;\text{d}g.
\end{align}

The probabilities for the encoded states and effects are:
\begin{align}
    \Tr(\rho_\inv E_\inv) = \int_G \int_G \Tr\left[ U_S(g) \rho_S U_S^\dagger(g) U_S(h) E_S U_S^\dagger(h) \right] \bra{\psi(g)}\gamma_R(h)\ket{\psi(g)} \;\text{d}g \;\text{d}h ,
\end{align}
we carry out the change of integration variable $g^{-1} h \to k$ and use the covariance of $\gamma_R$, $\bra{\psi(g)}\gamma_R(gk)\ket{\psi(g)} = \bra{\psi(e)}\gamma_R(k)\ket{\psi(e)}$, where $e\in G$ is the identity element:
\begin{align}
    &= \int_G\int_G \bra{\psi(e)}\gamma_R(k)\ket{\psi(e)} \; \Tr\left[ \rho_S U_S(k) E_S U_S^\dagger(k) \right] \;\text{d}g \;\text{d}k \label{eq:restriction} 
    = \Tr\left( \rho_S \; \int_G \text{d}k \; \mu(k) \, U_S(k) E_S  U_S^\dagger(k)  \right).
\end{align}
This is a trace of $\rho_S$ multiplying the effect $E_S$ averaged over all the coherent states of the system with their weights $\mu(k) = \bra{\psi(e)}\gamma_R(k)\ket{\psi(e)} = |\braket{\psi(e)|\tilde\psi(k)}|^2, \; \int_G \text{d}k \; \mu(k) = 1$, being the probability that the POVM $\gamma_R$ assigns $k$ to the coherent state corresponding to the identity element of the group. We note that in the case where the reference frame is ideal: $|\!\braket{\psi(g)|\psi(g')}\!|^2 = 0$ for $g \neq g'$, $F \propto \I$ and the weights reduce to $\delta(k)$, we obtain an exact encoding: $\Tr(\rho_S E_S) = \Tr(\rho_\inv E_\inv)$.

This construction is an instance of the relativisation/restriction framework of Ref.~\cite{Loveridge_2018}. Denote $U(g) = U_S(g)\otimes U_R(g)$, and the relativisation and restriction maps
\begin{align}
    \yen(A) := \int_G U_S(g)\,A\,U_S^\dagger(g)\otimes\gamma_R(g)\;\text{d}g, \qquad \Gamma_{\psi(e)}(A\otimes B) := A\,\bra{\psi(e)}B\ket{\psi(e)} .
\end{align}
Then
\begin{align}
    E_\inv = \yen(E_S), \qquad \rho_\inv = \int_G U(g)\big(\rho_S\otimes\ketbra{\psi(e)}{\psi(e)}\big)U^\dagger(g)\;\text{d}g, \qquad U(g)\,\yen(E_S)\,U^\dagger(g) = \yen(E_S),
\end{align}
and
\begin{align}
    \Tr(\rho_\inv E_\inv) = \Tr\big[\big(\rho_S\otimes\ketbra{\psi(e)}{\psi(e)}\big)\,\yen(E_S)\big] = \Tr\big[\rho_S\,\Gamma_{\psi(e)}\big(\yen(E_S)\big)\big] ,
\end{align}
which is \eqref{eq:restriction}. That is, the twirled state $\rho_\inv$ carries the expectation values of the invariant observables $\yen(E_S)$, \eqref{eq:restriction} is the restriction of the relativised effect to the reference state, and the weight $\mu(k)$ is the localisation of the reference state with respect to $\gamma_R$. For $\rho_S=\ketbra{\psi}{\psi}^\Sc$ and $\ket{\psi(e)}=\ket{G,G}^\Gc_z\ket{H,H}^\Hc_x$, $\rho_\inv$ is the incoherent average \eqref{eq:incoherent_average_sgh}. By \eqref{eq:restriction},
\begin{align}
    \big|\Tr(\rho_\inv E_\inv) - \Tr(\rho_S E_S)\big| \le \int_G\text{d}k\;\mu(k)\,\big\|U_S(k)E_SU_S^\dagger(k) - E_S\big\| ,
\end{align}
which vanishes as $\mu\to\delta_e$. The classical limit is thus the high-localisation limit of Ref.~\cite{Loveridge_2018}, reached along reference systems of growing dimension. Indeed, $\mu(k) = |\braket{\psi(e)|\tilde\psi(k)}|^2 = |\braket{\psi(k)| F^{-1/2} |\psi(e)}|^2$ and $F$ acts as a scaling on every total angular momentum $j$ sector. For $G=H\to\infty$, the reference state $\ket{\psi(e)}=\ket{G,G}^\Gc_z\ket{H,H}^\Hc_x$ has an approximately definite $j\approx \sqrt2 G$, meaning $F$ acts on it as scaling, and since $\braket{\psi(k) |\psi(e)} \to 0$ for $k\neq e$, $\mu$ must tend to the Dirac delta $\delta_e$.


\end{document}